\documentclass[letterpaper]{article} 
\usepackage{aaai2026}  
\nocopyright
\usepackage{times}  
\usepackage{helvet}  
\usepackage{courier}  
\usepackage[hyphens]{url}  
\usepackage{graphicx} 
\usepackage{natbib}  
\usepackage{caption} 
\usepackage{algorithm}
\usepackage{algorithmic}

\usepackage{newfloat}
\usepackage{listings}
\DeclareCaptionStyle{ruled}{labelfont=normalfont,labelsep=colon,strut=off} 
\floatstyle{ruled}
\newfloat{listing}{tb}{lst}{}
\floatname{listing}{Listing}
\usepackage{booktabs}
\usepackage{multirow}
\usepackage{amsmath}
\usepackage{amsfonts}

\usepackage{makecell}
\usepackage{tabularx}

\title{Automated Snippet-Alignment Data Augmentation for Code Translation}
\author{
    Zhiming Zhang,
    Qingfu Zhu,
    Xianzhen Luo,
    Yixuan Wang,
    Bohan Li,
    Wanxiang Che
}
\affiliations{
    \textsuperscript{\rm}Research Center for Social Computing and Interactive Robotics, \\ Harbin Institute of Technology, Harbin, China\\
    \{zmzhang, car\}@ir.hit.edu.cn
}

\usepackage{bibentry}

\begin{document}

\maketitle


\begin{abstract}
Code translation aims to translate the code from its source language to the target language and is used in various software development scenarios.
Recent developments in Large Language Models (LLMs) have showcased their capabilities in code translation, and parallel corpora play a crucial role in training models for code translation.
Parallel corpora can be categorized into program-alignment (PA) and snippet-alignment (SA) data.
Although PA data has complete context and is suitable for semantic alignment learning, it may not provide adequate fine-grained training signals due to its extended length,
while the brevity of SA data enables more fine-grained alignment learning.
Due to limited parallel corpora, researchers explore several augmentation methods for code translation.
Previous studies mainly focus on augmenting PA data.
In this paper, we propose a data augmentation method that leverages LLMs to generate SA data automatically.
To fully leverage both PA data and SA data, we explore a simple yet effective two-stage training strategy, which consistently enhances model performance compared to fine-tuning solely on PA data.
Experiments on TransCoder-test demonstrate that our augmented SA data combined with the two-stage training approach yields consistent improvements over the baseline, achieving a maximum gain of 3.78\% on pass@$k$. 

\end{abstract}

\begin{figure}[t]
  \centering
  \includegraphics[width=0.9\columnwidth]{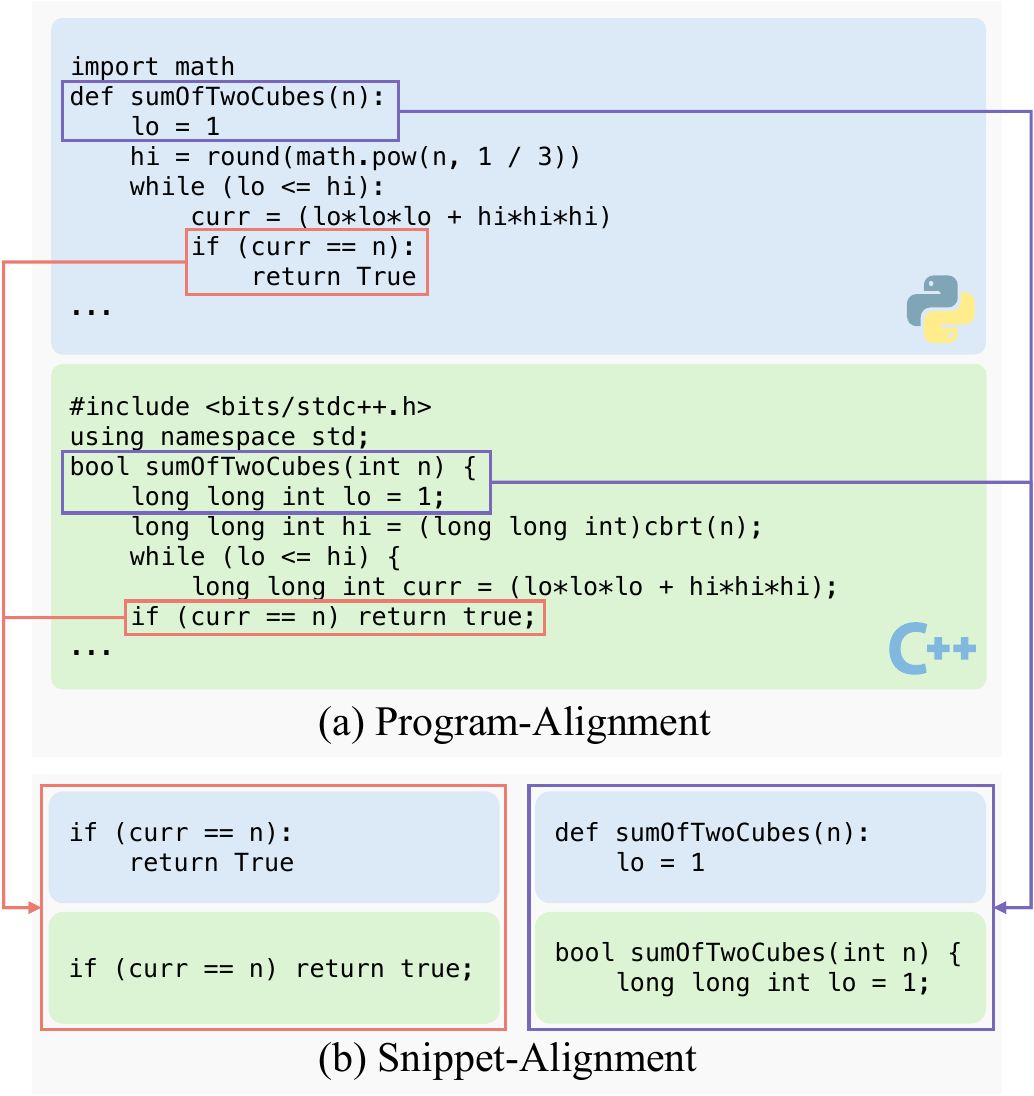}
  \caption{An example of PA data and SA data. 
  Although PA data is essential for learning semantic alignment due to its complete contextual information, it is usually lengthy.
  SA data, on the other hand, is brief and can provide more fine-grained training signals.}
  \label{fig:example}
\end{figure}
\section{Introduction}

Code translation aims to convert code between programming languages~\cite{chen2018tree}.
The advancement of automated code translation techniques has enhanced productivity in various scenarios,
such as migrating legacy software systems to modern programming languages~\cite{lachaux2020unsupervisedtranslation}, refactoring code bases~\cite{hong2023improvingc2rust,eniser2025translatingrealworldcodellms}, and enabling efficient cross-platform development~\cite{macedo2025intertrans}.
Early works have proposed some rule-based methods, including CxGO\footnote{CxGO: \url{https://github.com/gotranspile/cxgo}} and C2Rust\footnote{C2Rust: \url{https://github.com/immunant/c2rust}},
which are limited to specific language pairs and suffer from poor readability.
Recent advancements in Large Language Models (LLMs) have demonstrated their strong potential in code translation~\cite{pan2024lostintranslation,tao2024unraveling}.

Sufficient parallel corpora are essential for training code translation models.
From a data granularity perspective, existing parallel corpora can be divided into two categories: program-alignment (PA) and snippet-alignment (SA) data, as shown in Figure~\ref{fig:example}.
PA data consists of aligned program pairs that exhibit consistent behavior, and SA data follows a similar paradigm.
A program typically refers to a complete solution to a specific problem, and a snippet is a contiguous segment of a program~\cite{zhu2022xlcost}. 
While PA data provides complete contextual information for semantic alignment learning, which ensures behavioral consistency in translation, its lengthy sequence may make it challenging for models to focus on fine-grained alignment learning~\cite{pan2024lostintranslation,xinye2025enhancingllmslongcode}.
In contrast, the brevity of SA data facilitates the acquisition of 
more fine-grained alignment knowledge, such as syntactic alignment patterns~\cite{zhu2022xlcost}.
Both syntactic and semantic alignment learning are crucial for code translation. 
Therefore, both PA and SA data contribute to enhancing the code translation capability of the model.

The limited scenarios where parallel corpora can be acquired and the costly data acquisition constrain further improvements in model performance~\cite{chen2024data,li2024few}.
This motivates researchers to explore data augmentation methods for code translation.
Existing works primarily focus on augmenting PA data, including retrieval-based and LLM-based methods~\cite{xie-etal-2023-data,chen2024data}.
While PA data enables models to learn semantic alignment knowledge adequately, the lack of SA data may lead to syntactic errors or minor semantic errors during the translation~\cite{zhu2022xlcost, pan2024lostintranslation}.
For instance, a translated program may have no logic errors overall but exhibit behavioral inconsistencies due to incorrect API call translations~\cite{wang2025apirat}.
Conversely, SA data can complement PA data with finer training signals, enabling models to efficiently acquire fine-grained alignment knowledge, such as syntactic alignment knowledge.
Therefore, it is worthwhile to investigate augmenting SA data, which provides models with more fine-grained knowledge based on PA data.

In this paper, we propose a novel automated SA data augmentation pipeline, which bridges the gap in this area.
This pipeline takes PA data as input and leverages LLMs to generate SA data.
Concretely, the generation part of the pipeline operates in two stages: (1) leveraging LLMs to generate and insert comments into source programs, 
followed by (2) rewriting target programs that preserve both the content and order of the source-program comments, using the original target program as a reference. 
To effectively utilize both PA and SA data, we propose a training approach called 2-Stage, which employs sequential training to train the model first on one data granularity before transitioning to the other. 


We conduct comprehensive experiments on DeepSeek-Coder-Instruct and Qwen2.5-Coder-Instruct~\cite{guo2024deepseekcoder,hui2024qwen25}, which support various programming languages~\cite{he2025execoder}.
Experiments on TransCoder-test~\cite{lachaux2020unsupervisedtranslation} demonstrate that the model trained on augmented SA data combined with the proposed 2-Stage method can outperform the model trained solely on PA data, achieving a maximum gain of 3.78\% on pass@k.

Our contributions are summarized as follows:
\begin{itemize}
    \item We first analyze that existing augmentation methods can not provide sufficient fine-grained alignment knowledge, and we propose the first automated LLM-driven pipeline to generate SA data to address this issue.
    \item We propose a simple yet effective 2-Stage method that provides more fine-grained training signals, enabling more effective acquisition of alignment knowledge.
    \item Experimental results show that our model consistently outperforms baseline methods, highlighting the potential of SA data to further enhance model performance.
\end{itemize}

\begin{figure*}[t]
  \centering
  \includegraphics[width=0.9\textwidth]{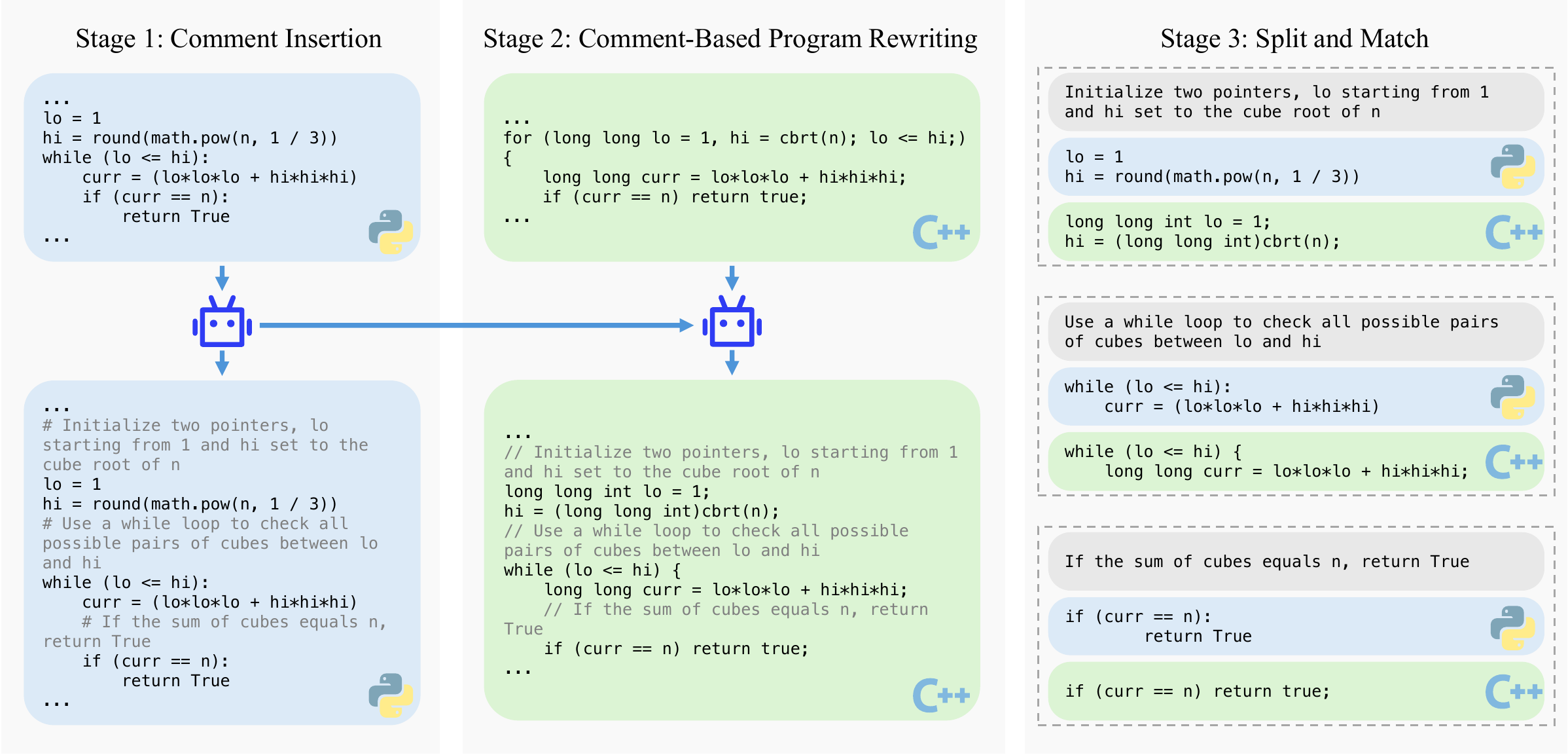}
  \caption{Overview of our data augmentation pipeline. It consists of two LLM-involved stages and a post-processing stage. 
  In \textbf{Stage 1} (the left part), the LLM takes the source program as input and outputs it with comments inserted.
  In \textbf{Stage 2} (the middle part), the LLM takes the output of Stage 1 and the original target program as input,
  then rewrites the target program to preserve the same content and order of comments in the source program.
  In \textbf{Stage 3} (the right part), we can split and match the code snippets according to the comments.}
  \label{fig:main_method}
\end{figure*}

\section{Related Work}
\subsection{Parallel Corpora in Code Translation} 

Scenarios where parallel corpora can be acquired are relatively limited, and human involvement is usually required.
For the PA dataset, they are primarily collected from programming websites~\cite{zhu2022xlcost,ahmad-etal-2023-avatar}, where solutions to the same problems in different languages naturally form parallel corpora.
Even on GitHub, which hosts abundant code repositories, parallel implementations of the same project in different programming languages remain relatively scarce~\cite{pan2024lostintranslation}.

For the SA dataset, CoST~\cite{zhu2022cost} is a multilingual code snippet translation dataset containing 7 commonly used programming languages.
All data in CoST was collected from GeeksForGeeks, which provided a standard template.
Through this template, the authors extracted the code snippets and comments for solutions in different languages, 
which were then aligned based on the comments.
Subsequently, XLCoST~\cite{zhu2022xlcost} significantly increased the amount of data available in CoST and extended its coverage to multiple programming tasks, including code translation, code summarization, and code synthesis.
Both datasets employ a comment-based segmentation method to derive SA data from PA data, though manual verification remains necessary in certain cases.


\subsection{Data Augmentation for Code Translation}
Due to the scarcity of parallel corpora in code translation, researchers have begun to explore data augmentation techniques.
\citet{xie-etal-2023-data} borrowed a term called \textit{comparable corpora} from natural language translation,
which referred to texts on similar topics in various languages~\cite{gete2022making}.
It also proposed a method of generating multiple references for the source code, which could provide more diverse training signals for the model.
\citet{zhu2024alignmentenhancing} first fine-tuned a small model on SA data~\cite{zhu2022xlcost}, 
then leveraged this model to generate parallel corpora from monolingual data, 
which significantly reduced reliance on parallel corpora.
\citet{chen2024data} designed a rule-based method to transform the data into a more diverse dataset.
It also proposed a method to retrieve similar source code and target code from a large code database to construct new parallel data.

However, existing data augmentation works mainly focus on augmenting PA data, 
and lack the exploration of SA data augmentation, which is essential for building a stronger code translation model.
In contrast, our work mainly focuses on augmenting SA data from the existing PA data, which bridges this gap in existing works.

\section{Automated SA Data Augmentation}

\subsection{Design Principles}
Before detailing the implementation details of our pipeline, we first outline its underlying design principles.
The foremost consideration in designing this pipeline is whether to generate data from scratch or to leverage existing PA data. 
For one thing, existing PA data can guarantee semantic correctness in program alignment~\cite{zhu2022xlcost,ahmad-etal-2023-avatar,yan-etal-2023-codetransocean,khan-etal-2024-xcodeeval}.
This allows us to focus primarily on establishing semantic alignment between code snippets. 
For another, generating data solely by LLM from scratch may lead to low diversity~\cite{wang-etal-2023-self-instruct,luo2024semiinstruct},
which could be even worse in SA data due to its inherent brevity.
Thus, we ultimately elect to generate SA data from existing PA data.

However, directly utilizing PA data to generate SA data still confronts two major challenges.
Firstly, it should be noted that different solutions corresponding to the same problem may use distinct algorithms or data structures, 
with variations in implementation details.
In other words, rewriting the target program may be necessary to ensure that the code snippets of the target program align with those of the source program.
Secondly, as the number of languages $N$ increases, generating SA data for all possible language pairs 
through a direct end-to-end generation requires $C(N, 2) = \frac{N(N-1)}{2}$ iterations,
which would significantly increase the computational resources needed.
For a certain problem, if we can align the segmentation patterns of all the other languages with that of a specific language, the number of required iterations will be reduced to $N$.

Therefore, we explicitly divide the generation part of the pipeline into two stages. 
The first stage will cache the snippet segmentation result to reduce computational resources.
In the second stage, the pipeline gives LLMs the flexibility to adaptively rewrite target programs as needed, 
thereby resolving internal snippet-alignment issues.

\subsection{Pipeline Implementation}
\label{sec:method}
Figure~\ref{fig:main_method} illustrates an overview of how to generate SA data from the original PA data.
Concretely, the proposed pipeline consists of two LLM-involved generation stages and a post-processing stage: 
(1) automatically generating and inserting code comments for source programs, 
followed by (2) rewriting target programs that preserve both the content and order of the source program comments 
using the original target program as a reference,
and (3) splitting the parallel comment-inserted programs and matching corresponding code snippets according to the comments.



\subsubsection{Comment Insertion}
\label{sec:comment_insertion}
Drawing inspiration from the construction methodology of XLCoST~\cite{zhu2022xlcost}, 
we use code comments as snippet separators.
%
Code comments, typically written in English, constitute a form of program annotation designed to enhance code readability and maintainability.
In contrast to traditional software engineering tools such as Abstract Syntax Tree (AST) and Control Flow Graph (CFG)~\cite{zhong-etal-2024-debug,du2025postincorporating}, 
comments are not language-specific and can take full advantage of the code comprehension capabilities of LLMs~\cite{cui2024codecomprehension}.

Given a source program $S$, we first prompt an LLM $M$ to generate and insert comments into $S$.
The insertion positions are entirely determined by $M$, 
guided solely by the principle of inserting as many comments as possible in order to ensure that each code snippet is not excessively lengthy.

After this stage, $M$ will output a new source program $S'= \left ( s_{0},c_{1},s_{1},c_{2},s_{2}...c_{k},s_{k}  \right ) $ 
containing $k$ comments and at most $k+1$ snippets, where $k$ is determined by $M$, $s_{i}$ denotes the $i$-th snippet of $S'$ and $c_{i}$ denotes the $i$-th comment of $S'$.
Note that $s_0$ can have a length of 0, since in some cases the first comment can be inserted at the very beginning. The lengths of the other snippets are non-zero.
\subsubsection{Comment-Based Program Rewriting}
\label{sec:rewriting}
Given a set of target programming languages $L = \left [ l_{1},l_{2}...l_{m}  \right ]$ containing $m$ kinds of languages 
and the original target program set $\tau = \left [ T_{1},T_{2}...T_{m}  \right ] $, where $l_{i}$ denotes the $i$-th target language and $T_{i}$ denotes the original target program written in $l_{i}$,
$M$ takes $S'$ and one original target program $T_{i}$ as input, and outputs the rewritten program $T'_{i}$.

Specifically, $M$ will rewrite $T_{i}$ based on the content and order of the comments in $S'$, using $T_{i}$ as a reference.
Notably, $T'_{i}$ may remain identical to $T_{i}$ when comments are disregarded, 
provided that $M$ believes that each snippet is already aligned and no rewriting process is required.

After this stage, for each $T_{i}$, $M$ will output a new target program $T'_{i}= \left ( t_{0}^{i},c_{1},t_{1}^{i},c_{2},t_{2}^{i}...c_{k},t_{k}^{i}  \right ) $ 
that theoretically containing exactly the same amount of comments and snippets as $S'$, where $t_{i}^{w}$ denotes the $i$-th snippet of $T_{w}$ and $c_{i}$ denotes the $i$-th comment of $T_{w}$.
Then we will obtain a new target program set $\tau' = \left [ T'_{1},T'_{2}...T'_{m}  \right ] $.

\subsubsection{Split and Match}
After the two stages above, we obtain $S'$ and $\tau'$ with identical comment content and comment order theoretically.
Suppose $S'$ as another target program $T'_{m+1}$, then we can formally construct a new target program set $\tau'' = \left [ T'_{1},T'_{2}...T'_{m},T'_{m+1}  \right ] $, where $T'_{m+1}=S'$.

Then for each pair of target programs $T'_{i}$ and $T'_{j}$, $i,j\in \left [ 1,m+1  \right ] $, we then extract each $t_{p}^{i}$ and $t_{p}^{j}$, $p\in \left [ 0,k  \right ] $, from $T'_{i}$ and $T'_{j}$ according to $c_{i}$ to construct the final snippet-alignment data.
If we find any discrepancy in the number or content of comments between $T'_{i}$ and $T'_{j}$, we will simply ignore this pair and continue the matching procedure.

\section{Experiment}
\subsection{Setting}

\paragraph{Datasets.}
We summarize the statistical information of the datasets used in our experiments in Table~\ref{tab:dataset_statistics}.
Following the previous works~\cite{pan2023stelo,he2025execoder},
we employ XLCoST~\cite{zhu2022xlcost} as the training set and the foundational dataset for our data augmentation,
which covers 7 commonly used programming languages and contains both PA and SA data.

We choose Python, Java, and C++ as our basic language setting, 
as they are widely used in different software development scenarios~\cite{lachaux2020unsupervisedtranslation, lachaux2021dobf, ahmad2023summarize, xinye2025enhancingllmslongcode},
and also cover both dynamic and static programming languages.
For the PA data, we select XLCoST-Program, i.e., 
the PA data in XLCoST, as the source of PA data used for PA training.
For the SA data, we employ the pipeline described in Section~\ref{sec:method} 
to construct an SA dataset, XLCoST-Augmented, 
using XLCoST-Program as the foundational dataset.
Specifically, we utilize DeepSeek-V3-0324 through its official API interface throughout the entire pipeline for Stage 1 and Stage 2.
We ultimately obtained 135654 pairs of SA data.

Following the previous works~\cite{yang2024exploring,du2025postincorporating,he2025execoder},
we have conducted a comprehensive set of experiments on TransCoder-test~\cite{lachaux2020unsupervisedtranslation}, 
a widely used public code translation dataset containing 1,788 pairs of PA data in Python, Java, and C++.
Each problem includes ten corresponding test cases to verify the semantic equivalence between the translated program and the source program, 
which enables more accurate measurement of the performance of LLMs in code translation, compared with those datasets without test cases~\cite{lu2021codexglue,zhu2022xlcost}.
\begin{table}[t]
\centering
\small
\begin{tabular}{lrl}
\toprule
\textbf{Dataset} & \textbf{Size (pairwise)} &\textbf{Granularity} \\
\cmidrule{1-3}

XLCoST-Program & 26972 & Program \\

XLCoST-Augmented & 135654 & Snippet \\

XLCoST-Snippet & 238875 & Snippet \\
\cmidrule{1-3}
TransCoder-test & 1788 & Program \\
\bottomrule
\end{tabular}
\caption{
Statistical information on training set and test set. 
The three upper rows represent the training set, while the bottom row corresponds to the test set.
}
\label{tab:dataset_statistics}
\end{table}
\paragraph{Evaluation Metric.}
Since we have access to the test cases of TransCoder-test, we can directly use the pass@$k$ metric 
to evaluate the semantic equivalence~\cite{chen2021evaluatinglargelanguagemodels}.
To calculate pass@$k$, we first generate $k$ code samples for each problem, 
and a problem is considered passed if at least one of the generated samples passes all the test cases.

Due to the issue of high variance brought by sampling, 
\citet{chen2021evaluatinglargelanguagemodels} refined it into a more stable metric as described in Equation~\ref{eq:passk}.
\begin{equation}
  \label{eq:passk}
  \text{pass@}k := \mathop{\mathbb{E}}_{\text{Problems}} \left[ 1 - \frac{{\binom{n-c}{k}}} {\binom{n}{k}} \right]
\end{equation}

For reproducibility purposes, we set $k=1$ and employ a greedy decoding strategy, which also reduces resource consumption during inference.

\begin{table*}[t]\footnotesize
\centering
\begin{tabularx}{\textwidth}{lccXXXXXXX}
\toprule
 & \textbf{Size} & \multicolumn{1}{c}{\textbf{Method}} & \textbf{J2P} & \textbf{C2P} & \textbf{P2C} & \textbf{J2C} & \textbf{P2J} & \textbf{C2J} & \textbf{AVG} \\
\cmidrule{1-10}

\multirow{6}{*}{\makecell{DeepSeek-Coder\\ Instruct}} 
& \multicolumn{1}{c}{\multirow{3}{*}{1.3B}}

& Vanilla 
& {65.91} & {71.93} & 64.95 & 82.46 & 65.58 & 78.15 & 71.50\\

& & Baseline (PP) & \textbf{83.12} & \textbf{82.75} & 77.62 & 87.24 & 83.12 & 88.36 & 83.70\\


& & 2-Stage-PS & \textbf{83.12} & 81.28 & \textbf{83.49} & \textbf{91.71} & \textbf{88.80} & \textbf{91.55} & \textbf{86.66}\\

\cmidrule{2-10}

& \multicolumn{1}{c}{\multirow{3}{*}{6.7B}}
& Vanilla 
& {76.95} & {81.65} & 56.88 & 63.96 & 81.66 & 86.12 & 74.54\\

& & Baseline (P) & \textbf{85.23} & \textbf{84.40} & 80.92 & 88.84 & 83.28 & 90.43 & 85.52\\

& & 2-Stage-PS & 84.58 & 83.30 & \textbf{88.81} & \textbf{93.78} & \textbf{91.07} & \textbf{94.26} & \textbf{89.30}\\

\cmidrule{1-10}

\multirow{6}{*}{\makecell{Qwen2.5-Coder\\ Instruct}} 
& \multicolumn{1}{c}{\multirow{3}{*}{1.5B}}

& Vanilla 
& {78.57} & {79.27} & 51.01 & 54.55 & 82.14 & 86.44 & 72.00\\

& & Baseline (PP)
& \textbf{84.74} & \textbf{84.95} & 77.98 & 87.72 & 82.79 & 87.56 & 84.29\\


& & 2-Stage-PS & 84.09 & 83.85 & \textbf{81.10} & \textbf{91.07} & \textbf{88.64} & \textbf{93.78} & \textbf{87.09}\\

\cmidrule{2-10}

& \multicolumn{1}{c}{\multirow{3}{*}{7B}}

& Vanilla 
& {83.60} & {82.39} & 74.31 & 72.25 & 88.64 & 88.04 & 81.54\\

& & Baseline (PP)
& {86.20} & {85.32} & 80.37 & 88.84 & 83.44 & 89.31 & 85.58\\

& & 2-Stage-PS & \textbf{86.85} & \textbf{86.42} & \textbf{85.32} & \textbf{93.14} & \textbf{89.45} & \textbf{93.78} & \textbf{89.16}\\

\bottomrule
\end{tabularx}
\caption{
Performance comparison of different methods.
The header \textbf{X2Y} denotes translating from language X to Y, 
where P, J, and C stand for Python, Java, and C++, respectively.
\textbf{AVG} stands for the average performance across all language pairs.
The \textbf{Vanilla} means evaluating the original model without any fine-tuning.
The \textbf{Baseline} denotes training the model solely on PA data, where the number of \textbf{P} in parentheses indicates the number of training epochs.
The \textbf{2-Stage-PS} means training the model sequentially on PA and SA data for one epoch respectively.
Our approach consistently outperforms the baseline on AVG score.
}
\label{tab:main_results}
\end{table*}

\paragraph{Training Details.}
For the backbone, we use DeepSeek-Coder-Instruct 1.3B/6.7B and Qwen2.5-Coder-Instruct 1.5B/7B, 
which are powerful open-source code LLMs supporting various programming languages~\cite{guo2024deepseekcoder,hui2024qwen25}.
Following the previous work~\cite{he2025execoder}, we employ instruction tuning on the backbone
using a simple code translation instruction template to better align its behaviour with code translation.

For the training hyperparameters, we set the batch size to 128 for PA data training and to 512 for SA data training.
We also set the max sequence length to 2048 for PA data training and to 1024 for SA data training.
For all the training procedures, we set the learning rate to 2e-5 and the warmup ratio to 0.1, using cosine lr\_scheduler\_type. 
We save the model once the last training epoch is completed.
For all the training, we use 2 NVIDIA A100-SXM4-80GB.

For the baseline, we directly fine-tune the model on PA data.
For our proposed 2-Stage method, we first train the model on data of one specific granularity, followed by training on data of another granularity.
We conduct all experiments twice and report the higher of the two results.

\subsection{Main Results}
\label{sec:main_results}
We present the performance comparison of our proposed 2-Stage method and the baseline in Table~\ref{tab:main_results}.
For the baseline, we train the model directly on PA data for one (represented by P) or two epochs (represented by PP), and report the best result.
The 2-Stage-PS here represents that we train the model on PA data for one epoch first, followed by training on SA data for one epoch.
Additionally, we report the result of the vanilla version of the backbone models, which are not trained on any PA or SA data.

Similar to the results of previous work~\cite{he2025execoder}, we can obtain a strong baseline model by fine-tuning solely on PA data, compared to the vanilla model.
This is typically the paradigm adopted when we want to adapt an LLM for code translation.
Furthermore, by applying our proposed 2-Stage method, we can consistently observe a significant average performance improvement across different sizes and backbone models, with metrics increasing by at least 2.8\% and up to 3.78\%.

This result is particularly intriguing because the models are ultimately trained on SA data yet evaluated on PA data, where a granularity mismatch exists.
From a curriculum learning perspective~\cite{bengio2009curriculum}, our 2-Stage approach essentially guides the model to first learn coarse-grained (PA) data before progressing to fine-grained (SA) data. This coarse-to-fine data organization may enable more comprehensive acquisition of alignment knowledge.

Moreover, the average performance gains primarily stem from X2J and X2C translations, whereas X2P translations fail to consistently outperform the baseline, with a slight degradation in some model settings.
This discrepancy may derive from differences between static programming languages (C++/Java) and dynamic programming languages (Python).
Static programming languages enforce stricter syntactic requirements, such as the explicit declaration of variable types, to enable errors to be detected during compilation.
In contrast, dynamic programming languages such as Python allow for runtime type modifications and have more flexible syntax rules.
Consequently, the fine-grained alignment knowledge learned from SA data may be more beneficial for X2C and X2J translations, while offering limited improvements for X2P translations.
Combined with the previously mentioned granularity mismatch between training and evaluation, these two factors may result in no significant improvement in X2P translations.

\begin{table*}[t]\footnotesize
\centering
\begin{tabularx}{\textwidth}{
   cc
   >{\centering\arraybackslash}X
   >{\centering\arraybackslash}X
   >{\centering\arraybackslash}X >{\centering\arraybackslash}X >{\centering\arraybackslash}X >{\centering\arraybackslash}X
   >{\centering\arraybackslash}X >{\centering\arraybackslash}X >{\centering\arraybackslash}X}
\toprule
 & \textbf{Size} & \textbf{Epoch} & \textbf{Order} & \textbf{J2P} & \textbf{C2P} & \textbf{P2C} & \textbf{J2C} & \textbf{P2J} & \textbf{C2J} & \textbf{AVG} \\
\cmidrule{1-11}

\multirow{14}{*}{\makecell{DeepSeek-Coder\\ Instruct}} 
& \multicolumn{1}{c}{\multirow{7}{*}{1.3B}}

& \multicolumn{1}{c}{\multirow{2}{*}{1}}

& P & \textbf{80.84} & \textbf{81.10} & 76.15 & 86.92 & 82.96 & 87.08 & 82.51\\
& & & S & 78.08 & 77.98 & \textbf{82.75} & \textbf{90.59} & \textbf{86.53} & \textbf{89.79} & \textbf{84.29}\\

\cmidrule{3-11}
& & \multirow{4}{*}{\makecell{2}}
& PP & \textbf{83.12} & \textbf{82.75} & 77.62 & 87.24 & 83.12 & 88.36 & 83.70\\
& & & SP & 82.79 & \textbf{82.75} & 77.25 & 86.92 & 83.44 & {88.20} & {83.56}\\
& & & SS & 80.20 & 80.18 & {82.20} & {89.95} & {86.53} & {90.43} & {84.92}\\
& & & PS & \textbf{83.12} & 81.28 & \textbf{83.49} & \textbf{91.71} & \textbf{88.80} & \textbf{91.55} & \textbf{86.66}\\

\cmidrule{2-11}
& \multicolumn{1}{c}{\multirow{7}{*}{6.7B}}

& \multicolumn{1}{c}{\multirow{2}{*}{1}}
& P & \textbf{85.23} & \textbf{84.40} & 80.92 & 88.84 & 83.28 & 90.43 & 85.52\\
& & & S & {83.77} & {82.94} & \textbf{87.52} & \textbf{92.66} & \textbf{91.07} & \textbf{94.26} & \textbf{88.70}\\

\cmidrule{3-11}
& & \multirow{4}{*}{\makecell{2}}
& PP & \textbf{86.69} & \textbf{86.06} & 78.90 & 89.31 & 78.25 & 89.00 & 84.70\\
& & & SP & 85.71 & {84.59} & 81.47 & 89.00 & 84.74 & {89.95} & {85.91}\\
& & & SS & 84.09 & 84.22 & {88.26} & {92.98} & \textbf{91.23} & {93.62} & {89.07}\\
& & & PS & 84.58 & 83.30 & \textbf{88.81} & \textbf{93.78} & {91.07} & \textbf{94.26} & \textbf{89.30}\\

\cmidrule{1-11}
\multirow{14}{*}{\makecell{Qwen2.5-Coder\\ Instruct}}
& \multicolumn{1}{c}{\multirow{7}{*}{1.5B}}
& \multicolumn{1}{c}{\multirow{2}{*}{1}}
& P & \textbf{84.74} & \textbf{84.40} & 74.86 & 86.60 & 81.98 & 88.68 & 83.54\\
& & & S & {83.28} & {82.50} & \textbf{80.92} & \textbf{91.71} & \textbf{86.36} & \textbf{93.14} & \textbf{86.32}\\

\cmidrule{3-11}
& & \multirow{4}{*}{\makecell{2}}
& PP & {84.74} & \textbf{84.95} & 77.98 & 87.72 & 82.79 & 87.56 & 84.29\\
& & & SP & \textbf{85.88} & \textbf{84.95} & 75.78 & 86.12 & 82.47 & {88.20} & {83.90}\\
& & & SS & 81.33 & 82.94 & \textbf{83.12} & {90.27} & {87.50} & {92.98} & {86.36}\\
& & & PS & 84.09 & 83.85 & {81.10} & \textbf{91.07} & \textbf{88.64} & \textbf{93.78} & \textbf{87.09}\\

\cmidrule{2-11}
& \multicolumn{1}{c}{\multirow{7}{*}{7B}}
& \multicolumn{1}{c}{\multirow{2}{*}{1}}
& P & \textbf{87.66} & \textbf{85.69} & 79.45 & 87.40 & 83.12 & 88.84 & 85.36\\
& & & S & {85.55} & {85.14} & \textbf{84.95} & \textbf{92.98} & \textbf{89.12} & \textbf{94.26} & \textbf{88.67}\\
\cmidrule{3-11}
& & \multirow{4}{*}{\makecell{2}}
& PP & {86.20} & {85.32} & 80.37 & 88.84 & 83.44 & 89.31 & 85.58\\
& & & SP & {86.04} & {84.40} & 79.63 & 87.56 & 84.25 & {89.00} & {85.15}\\
& & & SS & 85.23 & 82.75 & \textbf{85.87} & {92.34} & \textbf{89.61} & \textbf{94.74} & {88.42}\\
& & & PS & \textbf{86.85} & \textbf{86.42} & {85.32} & \textbf{93.14} & {89.45} & {93.78} & \textbf{89.16}\\

\bottomrule
\end{tabularx}
\caption{
Performance comparison of all order combinations of SA and PA data. \textbf{Epoch} represents the total training epochs. For example, \textbf{PP} represents training on PA data for 2 epochs, and \textbf{SP} represents training on SA data first for 1 epoch, followed by training on PA data for 1 epoch.
Among all combinations, \textbf{PS} consistently achieves the highest \textbf{AVG} score. 
}
\label{tab:combination_full}
\end{table*}

\subsection{Analysis}
\paragraph{Granularity Order of 2-Stage Training.}
To further investigate how the training order of different data granularities affects model performance, 
we conduct a series of experiments to evaluate the performance among different order combinations of data granularities.

Table~\ref{tab:combination_full} shows the performance comparison of all the order combinations.
When fixing the number of total training epochs to 1, 
we observe that training solely on SA data yields an average performance improvement of 2.76\% on AVG score, which is primarily attributed to improvements in the X2C and X2J translations.
A similar phenomenon can be observed when we extend the total training epochs to 2.
This aligns with our observations in Section \ref{sec:main_results}, 
where SA data is shown to be beneficial for X2C and X2J translations.

However, training solely on SA data fails to yield optimal results, 
necessitating the incorporation of PA data.
As a result, 2-Stage-PS consistently achieves the highest AVG score across all combinations,
illustrating that the training process should make full use of data of different granularities rather than relying on data of a single granularity.

In summary, training on single granularity fails to achieve optimal performance, whereas training on both PA and SA data yields the best results. 
Furthermore, the PS combination consistently outperforms other combinations, demonstrating the impact of granularity order on model performance.

\paragraph{SA Training as a Powerful Training Stage.}
From the results in Table~\ref{tab:combination_full}, 
we observe that training on SA data after PA data (2-Stage-PS) 
can yield a significant average performance improvement compared with training solely on PA data.
This leads us to investigate how model performance would vary if we add only one epoch of SA training 
after the PA training. 

As shown in Table~\ref{tab:epochs_full}, adding only one epoch of SA training after the PA training can consistently improve model performances for almost all the language pairs.
Besides, while the model exhibits overfitting as the number of PA training epochs increases, introducing SA training can continuously enhance the performance.
Moreover, even with the same total number of training epochs,
the model trained on both PA and SA data still significantly outperforms the model trained on PA data solely.


Overall, the results above demonstrate the promising potential of regarding SA training as a powerful training stage for achieving consistent model performance improvements.


\paragraph{Data Quality of LLM-Augmented SA Data.}

To further investigate the data quality of LLM-Augmented SA data, 
we present detailed statistical information of XLCoST-Augmented in Table~\ref{tab:augmented_statistics}.
After the augmentation, we obtain XLCoST-Augmented-Initial containing 139556 pairs of SA data.
We then post-process the SA data, including extracting the generated code from the output of the LLM,
comparing the content and order of comments of parallel data, and filtering out snippets that contain useless information (e.g., those consisting solely of \textit{import} or \textit{\#include} statements).

Finally, we discard 3902 pairs of SA data, and the overall data usability rate is 97.2\%,
demonstrating high-quality LLM-Augmented data and high effectiveness of our augmentation pipeline.

\begin{table}[t]
\centering
\small
\begin{tabular}{lrr}
\toprule
\textbf{Dataset} & \textbf{Size} & \textbf{Percentage}\\
\cmidrule{1-3}

XLCoST-Augmented-Initial & 139556 & 100\% \\

- Parsing Error & 24 & \textasciitilde0\% \\
- Comments Not Match Error & 425 & 0.31\% \\
- Filtering Error  & 3453 & 2.47\% \\
\cmidrule{1-3}
XLCoST-Augmented & 135654  & 97.20\%\\
\bottomrule
\end{tabular}
\caption{
Statistical information of data filtered by the entire post-processing stage.
}
\label{tab:augmented_statistics}
\end{table}

\begin{table*}[t]\footnotesize
\centering
\begin{tabularx}{\textwidth}{
   c
   c
   c
   >{\raggedright\arraybackslash}X
   >{\centering\arraybackslash}X >{\centering\arraybackslash}X >{\centering\arraybackslash}X >{\centering\arraybackslash}X
   >{\centering\arraybackslash}X >{\centering\arraybackslash}X >{\centering\arraybackslash}X}
\toprule
 & \textbf{Size} & \textbf{P-Epoch} & \textbf{Order} & \textbf{J2P} & \textbf{C2P} & \textbf{P2C} & \textbf{J2C} & \textbf{P2J} & \textbf{C2J} & \textbf{AVG} \\
\cmidrule{1-11}

\multirow{14}{*}{\makecell{DeepSeek-Coder\\ Instruct}}
& \multicolumn{1}{c}{\multirow{7}{*}{1.3B}}
& \multicolumn{1}{c}{\multirow{2}{*}{1}}
& P & {80.84} & {81.10} & 76.15 & {86.92} & 82.96 & 87.08 & 82.51\\

& & & PS & \textbf{83.12} & \textbf{81.28} & \textbf{83.49} & \textbf{91.71} & \textbf{88.80} & \textbf{91.55} & \textbf{86.66}\\

\cmidrule{3-11}
& & \multirow{2}{*}{\makecell{2}}
& PP & {83.12} & \textbf{82.75} & 77.62 & {87.24} & 83.12 & 88.36 & 83.70\\

& & & PPS & \textbf{84.09} & \textbf{82.75} & \textbf{81.65} & \textbf{91.07} & \textbf{88.64} & \textbf{92.50} & \textbf{86.78}\\

\cmidrule{3-11}
& & \multirow{2}{*}{\makecell{3}}
& PPP & {82.96} & {81.65} & 76.88 & {87.40} & 80.52 & 86.92 & 82.72\\

& & & PPPS & \textbf{84.42} & \textbf{82.57} & \textbf{84.59} & \textbf{91.71} & \textbf{89.61} & \textbf{92.98} & \textbf{87.65}\\

\cmidrule{2-11}
& \multicolumn{1}{c}{\multirow{7}{*}{6.7B}}
& \multicolumn{1}{c}{\multirow{2}{*}{1}}
& P & \textbf{85.23} & \textbf{84.40} & 80.92 & 88.84 & 83.28 & {90.43} & 85.52\\

& & & PS & 84.58 & 83.30 & \textbf{88.81} & \textbf{93.78} & \textbf{91.07} & \textbf{94.26} & \textbf{89.30}\\

\cmidrule{3-11}
& & \multirow{2}{*}{\makecell{2}}
& PP & \textbf{86.69} & \textbf{86.06} & 78.90 & {89.31} & 78.25 & 89.00 & 84.70\\

& & & PPS & {85.23} & {84.04} & \textbf{89.73} & \textbf{94.26} & \textbf{90.91} & \textbf{94.42} & \textbf{89.77}\\

\cmidrule{3-11}
& & \multirow{2}{*}{\makecell{3}}
& PPP & \textbf{85.55} & {83.49} & 74.68 & {88.52} & 72.24 & 86.92 & 81.90\\

& & & PPPS & {85.07} & \textbf{84.40} & \textbf{90.46} & \textbf{95.22} & \textbf{90.91} & \textbf{95.06} & \textbf{90.19}\\

\cmidrule{1-11}
\multirow{14}{*}{\makecell{Qwen2.5-Coder\\Instruct}}
& \multicolumn{1}{c}{\multirow{7}{*}{1.5B}}
& \multicolumn{1}{c}{\multirow{2}{*}{1}}
& P & \textbf{84.74} & \textbf{84.40} & 74.86 & 86.60 & 81.98 & {88.68} & 83.54\\

& & & PS & {84.09} & {83.85} & \textbf{81.10} & \textbf{91.07} & \textbf{88.64} & \textbf{93.78} & \textbf{87.09}\\

\cmidrule{3-11}
& & \multirow{2}{*}{\makecell{2}}
& PP & \textbf{84.74} & \textbf{84.95} & 77.98 & {87.72} & 82.79 & 87.56 & 84.29\\

& & & PPS & {83.93} & {84.04} & \textbf{82.75} & \textbf{91.87} & \textbf{88.80} & \textbf{92.34} & \textbf{87.29}\\

\cmidrule{3-11}
& & \multirow{2}{*}{\makecell{3}}
& PPP & \textbf{85.23} & \textbf{84.04} & 76.70 & {88.04} & 82.14 & 88.36 & 84.09\\

& & & PPPS & {83.93} & {83.49} & \textbf{84.04} & \textbf{91.55} & \textbf{87.50} & \textbf{93.94} & \textbf{87.41}\\

\cmidrule{2-11}
& \multicolumn{1}{c}{\multirow{7}{*}{7B}}
& \multicolumn{1}{c}{\multirow{2}{*}{1}}
& P & \textbf{87.66} & {85.69} & 79.45 & 87.40 & 83.12 & {88.84} & 85.36\\

& & & PS & {86.85} & \textbf{86.42} & \textbf{85.32} & \textbf{93.14} & \textbf{89.45} & \textbf{93.78} & \textbf{89.16}\\

\cmidrule{3-11}
& & \multirow{2}{*}{\makecell{2}}
& PP & \textbf{86.20} & {85.32} & 80.37 & {88.84} & 83.44 & 89.31 & 85.58\\

& & & PPS & \textbf{86.20} & \textbf{85.87} & \textbf{87.71} & \textbf{94.58} & \textbf{89.77} & \textbf{93.62} & \textbf{89.63}\\

\cmidrule{3-11}
& & \multirow{2}{*}{\makecell{3}}
& PPP & {85.88} & {85.14} & 79.08 & {88.52} & 82.63 & 89.16 & 85.07\\

& & & PPPS & \textbf{86.20} & \textbf{86.42} & \textbf{87.89} & \textbf{94.26} & \textbf{90.75} & \textbf{95.22} & \textbf{90.12}\\

\bottomrule
\end{tabularx}
\caption{
Performance comparison of adding only one epoch of SA training after the PA training.
\textbf{P-Epoch} denotes the number of PA training epochs.
As the P-Epoch varies, our models consistently outperform the model trained solely on PA data.
}
\label{tab:epochs_full}
\end{table*}

Furthermore, since XLCoST contains SA data, XLCoST-Snippet, which is manually constructed based on XLCoST-Program, we conduct experiments to evaluate the performance differences between models trained on our augmented SA data and the XLCoST-Snippet, using the identical 2-Stage-PS training methodology.
Table~\ref{tab:original_snippet} presents the performance comparison between the two SA datasets. 
We denote 2-Stage-PS$_{Ori}$ as the model trained on XLCoST-Snippet, and 2-Stage-PS$_{Aug}$ as its counterpart trained on XLCoST-Augmented.
Our experiments demonstrate that 2-Stage-PS$_{Aug}$ achieves significant superiority over 2-Stage-PS$_{Ori}$, proving that our augmented SA data surpasses the manually constructed XLCoST-Snippet in data quality.
\begin{table}[t]
\centering
\small
\begin{tabular}{lccc}
\toprule
 & \textbf{Size} & \multicolumn{1}{c}{\textbf{Method}} & \textbf{AVG} \\

\cmidrule{1-4}
\multirow{4}{*}{\makecell{DeepSeek-Coder\\ Instruct}} 
& \multicolumn{1}{c}{\multirow{2}{*}{1.3B}}

& 2-Stage-PS$_{Ori}$  & 83.27\\
& & 2-Stage-PS$_{Aug}$  & \textbf{86.66}\\

\cmidrule{2-4}
& \multicolumn{1}{c}{\multirow{2}{*}{6.7B}}

&  2-Stage-PS$_{Ori}$  & 85.81\\
& & 2-Stage-PS$_{Aug}$  & \textbf{89.30}\\

\cmidrule{1-4}
\multirow{4}{*}{\makecell{Qwen2.5-Coder\\ Instruct}} 
& \multicolumn{1}{c}{\multirow{2}{*}{1.5B}}

& 2-Stage-PS$_{Ori}$ & 83.23\\
& & 2-Stage-PS$_{Aug}$ & \textbf{87.09}\\

\cmidrule{2-4}

& \multicolumn{1}{c}{\multirow{2}{*}{7B}}

&  2-Stage-PS$_{Ori}$  & 85.17\\
& & 2-Stage-PS$_{Aug}$  & \textbf{89.16}\\
\bottomrule
\end{tabular}
\caption{
Performance comparison between model trained on original SA data or our augmented SA data, using 2-Stage-PS method.
}
\label{tab:original_snippet}
\end{table}
Notably, as shown in Table~\ref{tab:dataset_statistics}, XLCoST-Snippet contains almost twice the amount of SA data as XLCoST-Augmented, suggesting that the snippets in XLCoST-Snippet are generally shorter in length than those in XLCoST-Augmented.
We believe that excessively short snippets may corrupt its semantic integrity and potentially reduce data diversity.
In contrast, our augmentation approach leverages LLMs to generate comments for program segmentation, which better preserves the semantic integrity of individual snippets.
Moreover, we prompt LLMs to dynamically determine the number of generated snippets based on the length of the program, which ensures that the snippets remain of a moderate length.
And the moderate length actually can strike a balance between fine-grained alignment and data diversity.






\section{Conclusion}
\label{sec:conclusion}
One of the primary bottlenecks in code translation remains the scarcity of parallel corpora. 
These corpora can be divided into PA and SA data.
While PA data provides a complete context, its lengthy sequence may make it challenging for models to focus on fine-grained alignment learning.
Compared with PA data, the brevity of SA data enables it to provide more fine-grained training signals.
Prior works mainly focus on augmenting PA data, while overlooking the potential of SA data to provide additional fine-grained alignment knowledge.
In this paper, we propose an automated SA data augmentation method, 
which first bridges the gap in SA data augmentation.
Furthermore, we propose a straightforward yet effective 2-Stage method to fully leverage both PA and SA data during training.
Experimental results demonstrate that the model trained using the proposed 2-Stage method on our augmented data can consistently outperform the baseline, thus proving the effectiveness of both the augmented data and the proposed method.

\bibliography{aaai2026}

\end{document}